\documentclass[12pt]{article}
\input{psfig.sty}
\input{epsf.sty}
\usepackage{subfigure}
\usepackage{graphicx}
\usepackage{amsmath,amssymb}
\usepackage{color}
\usepackage{epsfig}
\newcommand{\Define}{\stackrel{\triangle}{=}}

\begin{document}
\baselineskip 0.25in

\title{\LARGE Full-Rate Full-Diversity Achieving MIMO Precoding with Partial
CSIT}
\author{
Biswajit Dutta, Somsubhra Barik and A. Chockalingam  \\ 
{\normalsize Department of ECE, Indian Institute of Science, 
Bangalore 560012, India }  \\
}
\date{}
\maketitle

\baselineskip 2.00pc

\begin{abstract}
In this paper, we consider a $n_t\times n_r$ multiple-input
multiple-output (MIMO) channel subjected to block fading. Reliability
(in terms of achieved diversity order) and rate (in number of symbols
transmitted per channel use) are of interest in such channels. We propose
a new precoding scheme which achieves both full diversity ($n_tn_r$th
order diversity) as well as full rate ($n_t$ symbols per channel use)
using partial channel state information at the transmitter (CSIT),
applicable in MIMO systems including $n_r<n_t$ asymmetric MIMO.
The proposed scheme achieves full diversity and improved coding gain
through an optimization over the choice of constellation sets. The
optimization maximizes $d_{min}^2$ for our precoding scheme subject to
an energy constraint. The scheme requires feedback of $n_t-1$ angle
parameter values, compared to $2n_tn_r$ real coefficients in case of
full CSIT. Error rate performance results for $3\times 1$, $3\times 2$,
$4\times 1$, $8\times 1$ precoded MIMO systems (with $n_t=3,3,4,8$
symbols per channel use, respectively) show that the proposed precoding 
achieves 3rd, 6th, 4th and 8th order diversities, respectively.
These performances are shown to be better than other precoding schemes 
in the literature; the better performance is due to the choice of the 
signal sets and the feedback angles in the proposed scheme.
\end{abstract}

\baselineskip 1.65pc
\newpage

\section{Introduction}
\label{sec1}
\vspace{-4mm}
Multiple-input multiple-output (MIMO) techniques can achieve high data
rates and spatial diversity in wireless communications over fading
channels \cite{paulraj},\cite{paul}. Spatial multiplexing (V-BLAST) with
$n_t$ antennas at the transmitter achieves the full rate of $n_t$ symbols
per channel use, but does not achieve transmit diversity. Space constraints
in user terminals like mobile/portable receivers make asymmetric MIMO
configuration with $n_r < n_t$ antennas at the receiver to be a preferred
choice. Precoding techniques can improve performance through the use of
channel state information at the transmitter (CSIT). Several precoding
methods use CSIT and achieve diversity benefits, but compromise on the
achieved rate; e.g., only one symbol per channel use is achieved in
transmit beamforming \cite{beamform}. It is desirable that precoding methods
achieve high rates (preferably the full rate of $n_t$ symbols per channel
use as in V-BLAST) and high diversity orders (preferably the full diversity
order of $n_tn_r$)\footnote{In this paper, `full rate' is defined as $n_t$
symbols per channel use regardless of the number of receive antennas, and
`full diversity' is defined as $n_tn_r$th diversity.} with partial CSIT.
Our work reported in this paper addresses this problem.

A vast body of research in the literature has addressed the problem of MIMO
precoding. In particular, precoding using partial CSIT/limited feedback has
been of interest because providing full CSIT (which refers to the full
knowledge of all the channel gains between transmit and receive antennas)
through feedback can be too expensive \cite{mukkavilli}-\cite{rohrao}.

Precoding in spatial multiplexing (V-BLAST) systems has been considered
for achieving high rates and transmit diversity \cite{collin}-\cite{BARIK}.
Precoding schemes in \cite{ASI}-\cite{sumeet} incur some loss in rate.
For example, the Grassmannian subspace packing based precoding in \cite{ASI}
does not allow simultaneous transmission of more than $n_t-1$ streams (i.e.,
achievable rate is $\leq n_t-1$ symbols per channel use). The precoding
scheme in \cite{KIM_POW} achieves full rate asymptotically\footnote{The 
scheme in \cite{KIM_POW} achieves a rate of $2T-1$ symbols per $T$ channel
uses in $2\times 2$ system, which achieves close to full rate (of $n_t=2)$
for large number of channel uses.} using D-BLAST
architecture with  partial CSIT, but achieves full diversity only for
$\min(n_t,n_r)=2$ and less diversity orders for other cases. Other recent
works have proposed to improve the diversity gain of singular value
decomposition (SVD) precoding\footnote{SVD precoding achieves the full 
rate of $n_t$ symbols per channel use but its diversity order is only 1.} 
by pairing good and bad subchannels, and jointly coding information across 
each pair of subchannels \cite{cross1},\cite{xycodes}. In the schemes in
\cite{cross1},\cite{xycodes},
though the achieved diversity orders are high, namely
$(n_t-\frac{n_s}{2}+1)(n_r-\frac{n_s}{2}+1)$ where $n_s \leq n_r$ is even 
number of streams, full diversity is not
guaranteed. Also, these schemes use full CSIT. In addition, the number
of symbols transmitted per channel use $n_s \leq n_r$, and so for
$n_r<n_t$ full rate is not achieved. The precoding scheme in \cite{BARIK}
uses partial CSIT in the form of a single angle parameter feedback. Though
this scheme achieves full rate and high diversity, it failed to achieve
full diversity. A summary of the required channel knowledge, achieved 
rates and diversity orders in various precoding schemes are given in 
Table-\ref{tab01}.

\begin{table}
\footnotesize
\label{tab01}
\begin{center}
\begin{tabular}{|c|c|c|c|}
\hline
Precoding & Channel & Rate & Diversity \\
& Knowledge & & \\ \hline \hline
Grassmannian  & Partial CSIT & Not full &  Full  \\ 
packing scheme \cite{beamform} & & $1$ symbol/chl. use & $n_tn_r$ \\ \hline
Grassmannian  & Partial CSIT & Not full &  Not Full  \\ 
packing scheme \cite{ASI} & & $ < n_t$ symbols/chl. use & \\ \hline
SVD precoding & Full CSIT & Not full &  Not full\\ 
& & $min(n_t,n_r)$ symbols/chl. use & 1 \\ \hline
E-$d_{min}$ Precoder in \cite{cross1}* & Full CSIT & Not full rate for $n_r<n_t$ & Not full \\
& & $n_s$ symbols/chl. use  & $(n_t-\frac{n_s}{2}+1)(n_r-\frac{n_s}{2}+1)$ \\
& & $n_s \leq n_r$, $n_t\geq n_r$, $n_s$ even & $n_s = \min(n_t,n_r)$, $n_s$ even \\ \hline
X-, Y-precoders in \cite{xycodes}* & Full CSIT & same as in \cite{cross1} & same as in \cite{cross1} \\ \hline 
Precoder in \cite{KIM_POW}** & Partial CSIT & Asymptotically full &  Not full for \\
& & $2T-1$ symbols in $T$ chl. uses &  $\min(n_t, n_r) \neq 2$ \\ 
& & for $n_t=n_r=2$ &  \\ \hline
Precoder in \cite{BARIK} & Partial CSIT & Full & Not full \\ 
& one angle & $n_t$ symbols/chl. use &  \\ \hline
Proposed & Partial CSIT & Full & Full \\ 
& $n_t-1$ angles & $n_t$ symbols/chl. use & $n_tn_r$ \\ \hline
\end{tabular}
\caption{{\footnotesize Comparison of required channel knowledge, achieved 
rates and diversity orders in different precoding schemes. \newline 
* E-$d_{min}$ and X-, Y-Codes in \cite{cross1} and \cite{xycodes}
differ mainly in complexity and applicability to higher-order QAM.
** Precoding scheme in \cite{KIM_POW} uses D-BLAST
architecture which needs multiple channel uses to asymptotically achieve 
full rate.}}
\end{center}
\end{table}

In the above context, the significance of our contribution in this paper
is that we propose a novel precoding scheme that achieves both {\em full
rate of $n_t$ symbols per channel use} as well as {\em full diversity of
$n_tn_r$} using {\em partial CSIT}. The scheme requires the feedback of
only $n_t-1$ angle parameter values, compared to $2n_tn_r$ real coefficients
in case of full CSIT. The proposed scheme achieves full diversity and coding
gain through an optimization over the choice of constellation sets. The
optimization maximizes $d_{min}^2$ for our
precoding scheme subject to an energy constraint. We present codeword
error performance results for $n_t=3,4,8$ and $n_r=1,2$. We analytically 
prove the full diversity ($n_tn_r$th diversity) of the proposed scheme, 
and simulation results are shown to validate the result. It is further shown 
that the proposed scheme performs better than other existing schemes. This 
better performance is attributed to the choice of signal sets and feedback 
angles in the proposed scheme.

{\em Notation:} Vectors are denoted by lower case boldface letters, and
matrices are denoted by upper case boldface letters. $(.)^*$, $(.)^H$,
$(.)^T$ and $\mbox{tr}(.)$ denote conjugation, hermitian, transpose and
trace operators, respectively. ${\bf A}_{xy}$ will be used to denote the
$(x,y)$th entry of matrix ${\bf A} $.

The rest of the paper is organized as follows. The system model is
presented in Section \ref{sec2}. The proposed full rate, full diversity
precoding scheme, and choice of constellation sets are presented in 
Section \ref{sec3}. Performance results and discussions are presented 
in Section \ref{simu}. Conclusions are given in Section \ref{sec6}.

\vspace{-4mm}
\section{System Model }
\vspace{-4mm}
\label{sec2}
Consider a precoded V-BLAST system with $n_t$ antennas at the transmitter
and $n_r$ antennas at the receiver. Let
${\bf H}\in \mathbb{C}^{n_r\times n_t}$ denote the channel gain matrix,
whose entries $h_{pq}$ are distributed i.i.d $\mathcal{CN}(0,1)$,
$\forall p=1, \cdots, n_r, \ \forall q = 1,\cdots, n_t$. Let ${\bf F}$
denote the precoder matrix of size $n_t\times n_t$, which is known to
both the transmitter and receiver. For a given channel matrix ${\bf H}$,
the receiver computes $n_t-1$ number of real feedback parameters and
sends them to the transmitter. Given the feedback parameters, the
transmitter forms the precoding matrix ${\bf F}$. Let ${\bf x}$ denote
the $n_t \times 1 $ complex data symbol vector of the form
\begin{equation}
\begin{array}{cc}
{\bf x} \\
\end{array} = \left[\begin{array}{ccc}
x_1 \,\, x_2 & \cdots & x_{n_t}  \\
\end{array}\right]^T,
\end{equation}
where $x_i$ is transmitted by the $i$th antenna. $x_i$'s will take values
from a suitable complex constellation, the choice of which will be discussed
later. The received signal
vector, ${\bf y} \in \mathbb C^{n_r\times 1}$, at the receiver is given by
\begin{eqnarray}
{\bf y} & = & {\bf H}{\bf F}{\bf x} + {\bf n},
\label{eqn1}
\end{eqnarray}
where ${\bf n} \in \mathbb C^{n_r\times 1}$ is the noise vector with its
entries distributed as  i.i.d. $\mathcal C \mathcal N\big(0,\sigma^2)$.

\vspace{-4mm}
\section{Proposed Precoding Scheme}
\label{sec3}
\vspace{-4mm}
Consider the precoded V-BLAST system described in Section \ref{sec2}.  Let
\begin{equation}
\begin{array}{cc}
{\bf F} \\
\end{array} = \left[\begin{array}{ccc}
{\bf a} \,\,\, {\bf a} & \cdots & {\bf a}  \\
\end{array}\right],
\label{eq6}
\end{equation}
where the column vector
${\bf a}\ =\ [a_1 \ \cdots \ a_{n_t}]^T \ \in \mathbb{C}^{n_t \times 1}$
and $ |a_i| = 1, \forall i=1,\cdots,n_t$. At high SNRs, symbol error
probability $P_e$ decays as SNR$^{-n_tn_r}$, provided the received distance
squared between each pair of codewords ${\bf x}_k, {\bf x}_l$ $( k \neq l)$,
denoted by ${d_{k,l}^2}$, is a Chi-square $\chi_{2n_tn_r}^2$ distributed
variable with $2n_tn_r$ degrees of freedom \cite{BARIK}. In general,
\begin{eqnarray}
\label{dmin_norm_form}
d_{k,l}^2 &=& \|{\bf HF}\Delta {\bf x}\|^2,
\end{eqnarray}
where $\Delta{\bf x}={\bf x}_k-{\bf x}_l$, $k \neq l$.
Now, (\ref{dmin_norm_form}) can be written as
\begin{eqnarray}
d_{k,l}^2 &=& \mbox{tr}({\bf HF}\Delta {\bf x}\Delta {\bf x}^H{\bf F}^H{\bf H}^H) \nonumber \\
& = & \mbox{tr}({\bf F}^H{\bf H}^H{\bf HF}\Delta{\bf  x}\Delta {\bf x}^H).
\label{dmin_trace2}
\end{eqnarray}
It can be shown that the $(p,q)$th element of ${\bf H}^H{\bf HF}$ is given by
\begin{eqnarray}
({{\bf H}^H{\bf HF}})_{pq} & = & \sum_{m=1}^{n_t}a_m\Big(\sum_{o=1}^{n_r}{h}^*_{op}h_{om}\Big).
\end{eqnarray}
It may be noted that all the elements of the $p$th row of the matrix
${\bf H}^H{\bf HF}$ are identical.
\begin{eqnarray}
\Rightarrow({\bf F}^H{\bf H}^H{\bf HF})_{pq}& = & \sum_{n=1}^{n_t}{a}^*_n\Big(\sum_{m=1}^{n_t}a_m\Big(\sum_{o=1}^{n_r}{h}^*_{on}h_{om}\Big)\Big) \nonumber \\
& = & \sum_{n=m=1}^{n_t}|a_{n}|^2\Big(\sum_{o=1}^{n_r}|h_{on}|^2\Big) 
+ \, \sum_{n\neq m}^{n_t}{a}^*_na_m\Big(\sum_{o=1}^{n_r}{h}^*_{on}h_{om}\Big).
\end{eqnarray}
Moreover, it is clear that all the entries of the matrix
${\bf F}^H{\bf H}^H{\bf HF}$ are identical. So it amounts to 
\begin{eqnarray}
{\bf F}^H{\bf H}^H{\bf HF}& = & \left( \sum_{n=m=1}^{n_t}|a_{n}|^2\Big(\sum_{o=1}^{n_r}|h_{on}|^2\Big) 
+ \, 2\mathcal{R} \Big(\sum_{m,n,n>m}^{n_t}{a}^*_na_m\sum_{o=1}^{n_r}{h}^*_{on}h_{om}\Big)\right)\bf{B},
\label{clever_expr}
\end{eqnarray}
where ${\bf B}_{pq}=1$. In order to ensure that $d_{k,l}^2$ is distributed
as a Chi-square $\chi_{2n_tn_r}^2$ variable, we require that in
(\ref{clever_expr}),
\begin{eqnarray}
\mathcal{R}\Big(\sum_{m,n,n>m}^{n_t}{a}^*_na_m\Big(\sum_{o=1}^{n_r}{h}^*_{on}h_{om}\Big)\Big)&=& 0.
\label{zero_eqn}
\end{eqnarray}
Let $a_i=|a_i|e^{j\theta_i}$, $\sum_{o=1}^{n_r}{h}^*_{on}h_{om}=|\sum_{o=1}^{n_r}{h}^*_{on}h_{om}|e^{j\alpha_{nm}}$,
where $j=\sqrt{-1}$. Then, with $|a_i|=1$, $\forall i=1,\cdots,n_t$,
(\ref{zero_eqn}) becomes
\begin{equation}
\sum_{m,n,{n>m}}^{n_t}\big|\sum_{o=1}^{n_r}{h}^*_{on}h_{om}\big|\cos(\theta_m-\theta_n+\alpha_{nm})\ =\ 0.
\label{eqn9}
\end{equation}
For a given realization of $\bf H$, the receiver needs to compute
$\theta_m$'s to satisfy (\ref{eqn9}). Again, (\ref{eqn9}) can be rewritten as
\begin{eqnarray}
\label{inner_sum}
\sum_{n=2}^{n_t}\sum_{m=1}^{n-1}\big|\sum_{o=1}^{n_r}{h}^*_{on}h_{om}\big|\cos(\theta_m-\theta_n+\alpha_{nm})& \hspace{-1mm} =& \hspace{-1mm} 0.
\end{eqnarray}
Let the inner summation of (\ref{inner_sum}) with index $m$ be equal to 0
for each $n$. For $n=2$,
\begin{eqnarray}
\cos(\theta_1-\theta_2+\alpha_{21}) & = & 0.
\end{eqnarray}
Let $\theta_1=0$. Hence, $\theta_2=\alpha_{21} -\pi/2$.
Next, for any $n>2$,
\begin{eqnarray}
\sum_{m=1}^{n-1}\big|\sum_{o=1}^{n_r}{h}^*_{on}h_{om}\big|\cos(\theta_m-\theta_n+\alpha_{nm})& = &  0 \\ \nonumber
\Rightarrow \quad \sum_{m=1}^{n-1}\big|\sum_{o=1}^{n_r}{h}^*_{on}h_{om}\big|(\cos(\theta_m+\alpha_{nm})\cos \theta_n 
+ \sin(\theta_m+\alpha_{nm})\sin \theta_n)& = & 0. \nonumber
\end{eqnarray}
\begin{eqnarray}
\therefore \quad \theta_n & = & \tan^{-1}\left(\frac{-\sum\limits_{m=1}^{n-1}|\sum_{o=1}^{n_r}{h}^*_{on}h_{om}|(\cos(\theta_m+\alpha_{nm})}{\sum\limits_{m=1}^{n-1}|\sum_{o=1}^{n_r}{h}^*_{on}h_{om}|\sin(\theta_m+\alpha_{nm})}\right).
\label{recur_formul}
\end{eqnarray}
Using (\ref{recur_formul}), $\theta_n$ can be recursively calculated,
given the values of  $\theta_1$ to $\theta_{n-1}$. From
(\ref{dmin_trace2}),(\ref{clever_expr}),(\ref{zero_eqn}) and
(\ref{recur_formul}), after simplification, we have
\begin{eqnarray}
d_{k,l}^2 & = & \Big(\sum_{n=1}^{n_t}\sum_{o=1}^{n_r}|h_{on}|^2\Big)\big|\sum_{i=1}^{n_t}\Delta x_i\big|^2,
\label{dmin_final_expr}
\end{eqnarray}
where
\begin{equation}
\begin{array}{cc}
\Delta{\bf x} \\
\end{array} \, = \, \left[\begin{array}{ccc}
\Delta x_1 \,\,\, \Delta x_2 & \cdots &  \Delta x_{n_t}
\end{array}\right]^{T}.
\label{eqn12}
\end{equation}

\subsection{Choice of Constellation Sets}
\label{const}
\vspace{-2mm}
From (\ref{dmin_final_expr}), it is clear that full diversity is guaranteed
if, for each codeword pair ${\bf x}_k,{\bf x}_l$, $k\neq l$ in the codebook,
\begin{eqnarray}
\label{condition}
\sum_{i=1}^{n_t}\Delta x_i & \neq & 0.
\end{eqnarray}
Let $x_i$, $\forall i=1,\cdots,n_t$, take values from some set
$C_i\subseteq $ $a_i \mathbb{Z}[j]=\{ a_i z:z\in \mathbb{Z}[j]\}$,
where $\mathbb{Z}[j]=u+jv$, $u$ and $v$ are integers, i.e., regular
QAM constellations scaled by $a_{i}$. Clearly then, $\Delta x_i$ belongs
to some other set $D_i\subseteq a_{i}\mathbb{Z}[j]$ for each $i$. It may
be noted that $-\Delta x_i \in D_i$. Now, let $a_1=1$ and choose $a_2$
such that
\begin{eqnarray}
D_1\cap D_2 & = & \{0\}.
\end{eqnarray}
Again, define $D_1+D_2\Define \{\Delta x_1+\Delta x_2 \,\, \mbox{s.t}\,\, \forall \Delta x_1 \in D_1, \Delta x_2 \in D_2\}$.
Next, choose $a_3$ such that
\begin{eqnarray}
D_3\cap (D_1+D_2) & = & \{0\}.
\end{eqnarray}
Proceeding in this way, $a_i$ is chosen such that
\begin{eqnarray}
\label{inter}
D_i\cap (D_1+\cdots + D_{i-1}) & = & \{0\}.
\end{eqnarray}
This implies that for such choice of $a_i$'s and
$\Delta x_i \in D_i$,
(\ref{condition}) is satisfied.

A trivial choice of $a_{i}$'s
could be $\sqrt{p_i}$, where $p_i$'s are distinct prime numbers in
$\mathbb{Z}$. This shows that the solution set of $a_i$'s satisfying
(\ref{inter}) is non-empty. For these values of $a_i$'s, (\ref{condition})
holds true for all codeword pairs in the codebook ensuring full diversity.
Since the number of elements of $C_i$ is finite, the cardinality of $D_i$
is also finite. This enables us to exhaustively enumerate all the elements
of $D_i$ as functions of $a_i$. Next, we have to choose
$a_i$'s $\in {\mathbb C}$ such that (\ref{inter}) is satisfied
$\forall i=1,\cdots,n_t$. Thus, there exist scaled QAM constellations for
which the proposed precoding scheme guarantees full diversity. A parallel
argument can be made for PSK constellation to ensure full diversity.

Now, the $d_{min}^2$ parameter of a codebook determines its coding gain as
given by the error probability expression at high SNR. Furthermore,
$d_{min}^2$ is maximized if $min(\sum_{i=1}^{n_t}\Delta x_i)$ is maximized
in (\ref{dmin_final_expr}). Hence, after scaling the constellation sets
$C_i$'s by appropriate $a_{i}$'s, we propose to rotate/scale each $C_i$ by
angles ($\phi_{i}$'s) and scaling real numbers ($b_{i}$'s), ensuring that
the average constellation power never exceeds the total transmit power
constraint and also the condition (\ref{inter}) holds true
$\forall i=1,\cdots,n_t$. The corresponding
$\min(\sum_{i=1}^{n_t}\Delta x_i)$ is computed over all $C_i$'s. The
optimum rotation angle ($\phi_{i,opt}$) and scaling real number
($b_{i,opt}$) for each $C_i$ are chosen by computer search for which
$\min(\Sigma_{i=1}^{n_t}\Delta x_i)$ is maximum. This fixes the choice
of constellation sets $C_i$'s that guarantee full transmit diversity
as well maximizes $d_{min}^2$ subject to transmit power constraint.

\subsection{A Simplified Approach to Choose Constellation Sets}
\label{simple}
\vspace{-2mm}
The treatment described in the previous subsection (Section {\ref{const}})
is feasible for constellations of small cardinality. The computational
complexity involved in choosing large-sized constellation sets makes
the choice cumbersome. In this subsection, we illustrate a suboptimal
mechanism to choose large constellations. Our choice of ${\bf F}$ in
(\ref{eq6}) ensures that the effective transmitted vector ${\bf Fx}$
becomes
\begin{eqnarray}
\Big(\sum_{i=1}^{n_t}x_i\Big)\left[\begin{array}{ccc}
a_1 \,\,\, a_2 & \cdots &  a_{n_t}  \\
\end{array}\right]^{T}.
\end{eqnarray}
It then implies that each of the $n_t$ antennas effectively transmits the
symbol $\sum_{i=1}^{n_t}x_i$. Now, for a given energy constraint, it is
known that square QAM constellations (i.e., $4^f$ sized constellation sets,
where $f$ is a positive integer) achieve better $d_{min}^2$ over PAM and
PSK constellations. Hence, we propose to choose constellation sets for
each $x_{i}$ in such a way that $\sum_{i=1}^{n_t}x_i$ takes values from
QAM constellation. Such a choice guarantees full diversity since for each
codeword ${\bf x}$ transmitted the corresponding effective symbol
$\sum_{i=1}^{n_t}x_i$ maps to a unique point in some QAM constellation
satisfying condition (\ref{condition}).

\subsection{Constellation Sets Obtained for $n_t=3,4,8,16$}
\label{constellation}
\begin{figure}[h]
\hspace{1cm}
\centering
\subfigure[1 bit/symbol, $n_t=4$]{
\includegraphics[width=2.80in,height=2.5in]{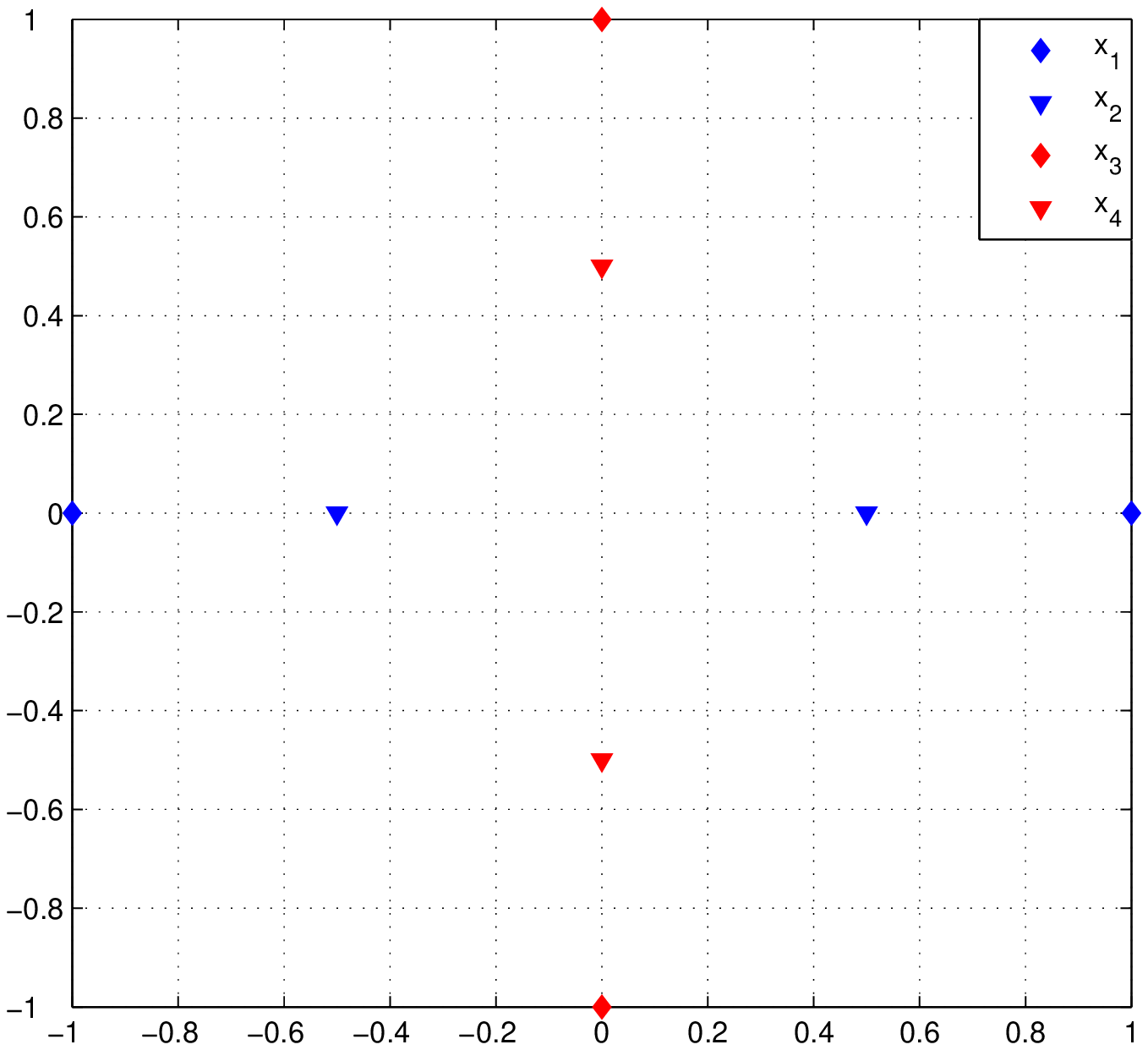}}
\hspace{3mm}
\subfigure[2 bits/symbol, $n_t=4$]{
\includegraphics[width=2.80in,height=2.5in]{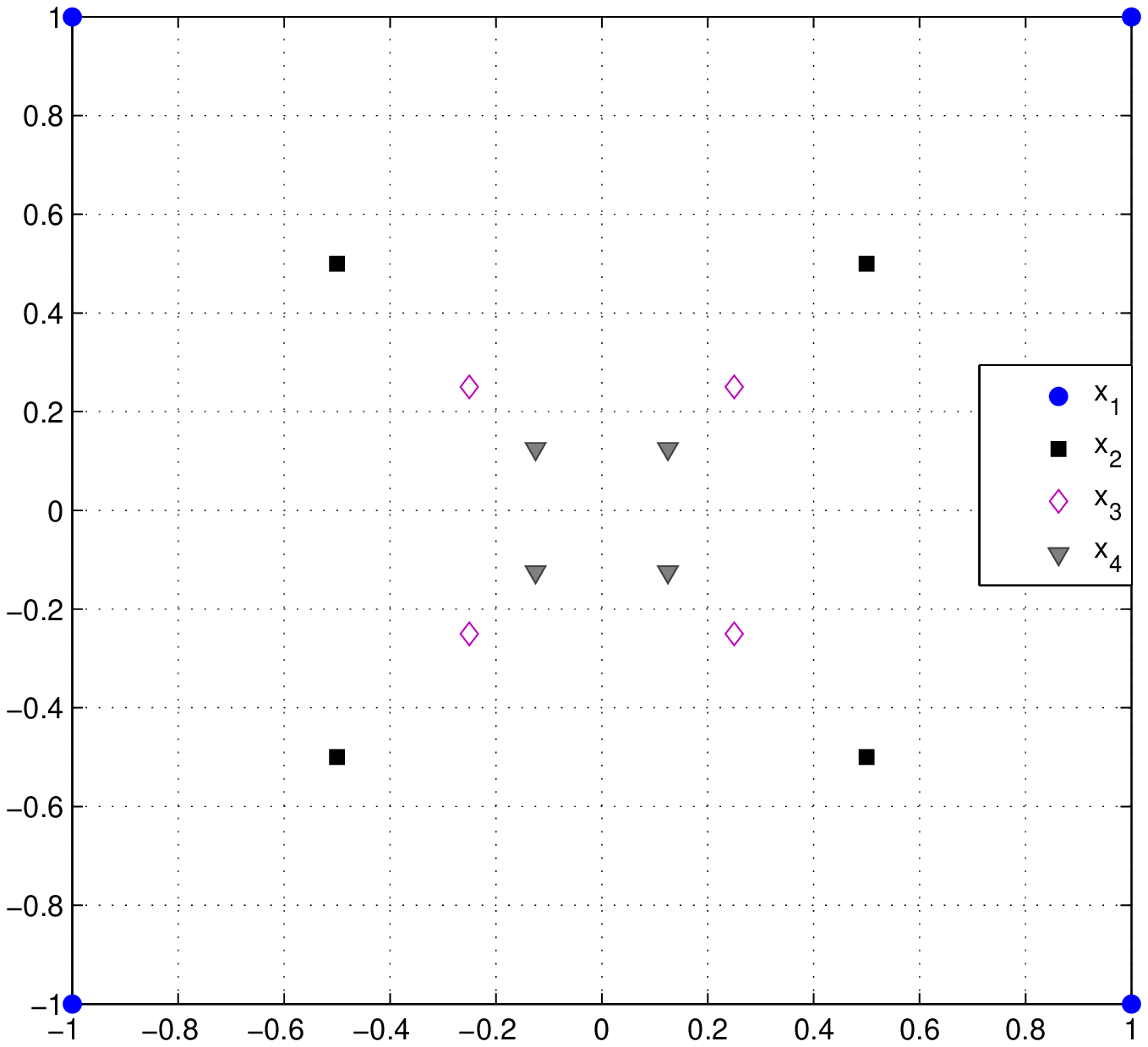}}
\vspace{-2mm}
\caption{Plots of constellation sets for $n_t=4$ with 1 bit/symbol
and 2 bits/symbol.}
\label{const1}
\end{figure}

\vspace{-2mm}
Based on the treatment described in the above subsections (Sections
\ref{const}, \ref{simple}), we carried out a computer search to obtain
full diversity achieving constellation sets for various $n_t$. Let 
$a {\cal Q}_M=\{a q : q\in {\cal Q}_M,a \in \mathbb{R}\}$, where
${\cal Q}_M $ denotes $M$-QAM constellation. The obtained constellation 
sets for $n_t=3,4,8,16$ are given in Table-2. Figures \ref{const1}(a) 
and \ref{const1}(b) show the plots of the obtained constellation sets
for $n_t=4$ with of 1 bit/symbol and 2 bits/symbol, respectively.

\begin{table}
\label{tab11}
\footnotesize 
\begin{center}
\begin{tabular}{|c|c|c|c|}
\hline
No. of Tx antennas  & Modulation & Modulation & Modulation \\
 & (1 bit/symbol) & (2 bits/symbol) & (4 bits/symbol) \\ \hline
$n_t=3$ & $x_1\in\{\pm 1\}$ & $x_1\in{\cal Q}_4$ & $ x_1 \in {\cal Q}_{16} $\\
& $x_2\in\{\pm j\}$& $x_2\in\frac{1}{2}{\cal Q}_4$ & $ x_2 \in \frac{1}{14} {\cal Q}_{16} $ \\
& $x_3\in\{\pm 0.675e^{j\frac{\pi}{4}}\}$& $x_3\in\frac{1}{4}{\cal Q}_4$ & $ x_3 \in \frac{1}{28} {\cal Q}_{16} $\\ \hline
$n_t=4$ & $x_1\in\{\pm 1\}$ & $x_1\in{\cal Q}_4$ & $ x_1 \in {\cal Q}_{16} $ \\
& $x_2\in\{\pm j\}$& $x_2\in\frac{1}{2}{\cal Q}_4$ & $ x_2 \in \frac{1}{14} {\cal Q}_{16} $\\
& $x_3\in\{\pm \frac{1}{2} \}$& $x_3\in\frac{1}{4}{\cal Q}_4$ & $ x_3 \in \frac{1}{28} {\cal Q}_{16} $\\
& $x_4\in\{\pm \frac{1}{2}j \}$& $x_4\in\frac{1}{8}{\cal Q}_4$ & $ x_4 \in \frac{1}{56} {\cal Q}_{16} $\\ \hline
$n_t=8$ & $x_1\in\{\pm 1\}$ & $x_1\in{\cal Q}_4$ & $ x_1 \in {\cal Q}_{16} $\\
& $x_2\in\{\pm j\}$& $x_2\in\frac{1}{2}{\cal Q}_4$ & $ x_2 \in \frac{1}{14} {\cal Q}_{16} $\\
& $x_3\in\{\pm \frac{1}{2} \}$& $x_3\in\frac{1}{4}{\cal Q}_4$ & $ x_3 \in \frac{1}{28} {\cal Q}_{16} $ \\
& $x_4\in\{\pm \frac{1}{2}j \}$& $x_4\in\frac{1}{8}{\cal Q}_4$ & $ x_4 \in \frac{1}{56} {\cal Q}_{16}$\\
& $x_5\in\{\pm \frac{1}{4} \}$& $x_5\in\frac{1}{16}{\cal Q}_4$ & $ x_5 \in \frac{1}{112} {\cal Q}_{16}$\\
& $x_6\in\{\pm \frac{1}{4}j \}$& $x_6\in\frac{1}{32}{\cal Q}_4$ & $ x_6 \in \frac{1}{224} {\cal Q}_{16}$ \\
& $x_7\in\{\pm \frac{1}{8} \}$& $x_7\in\frac{1}{64}{\cal Q}_4$ & $ x_7 \in \frac{1}{448} {\cal Q}_{16}$\\
& $x_8\in\{\pm \frac{1}{8}j \}$& $x_8\in\frac{1}{128}{\cal Q}_4$ & $ x_8 \in \frac{1}{896} {\cal Q}_{16}$\\ \hline
$n_t=16$ & $x_1\in\{\pm 1\}$ & $x_1\in{\cal Q}_4$ & $ x_1 \in {\cal Q}_{16} $\\
& $x_2\in\{\pm j\}$& $x_2\in\frac{1}{2}{\cal Q}_4$ & $ x_2 \in \frac{1}{14} {\cal Q}_{16} $\\
& $x_3\in\{\pm \frac{1}{2} \}$& $x_3\in\frac{1}{4}{\cal Q}_4$ & $ x_3 \in \frac{1}{28} {\cal Q}_{16} $\\
& $x_4\in\{\pm \frac{1}{2}j \}$& $x_4\in\frac{1}{8}{\cal Q}_4$ & $ x_4 \in \frac{1}{56} {\cal Q}_{16}$\\
& $x_5\in\{\pm \frac{1}{4} \}$& $x_5\in\frac{1}{16}{\cal Q}_4$ & $ x_5 \in \frac{1}{112} {\cal Q}_{16}$\\
& $x_6\in\{\pm \frac{1}{4}j \}$& $x_6\in\frac{1}{32}{\cal Q}_4$ & $ x_6 \in \frac{1}{224} {\cal Q}_{16}$\\
& $x_7\in\{\pm \frac{1}{8} \}$& $x_7\in\frac{1}{64}{\cal Q}_4$ & $ x_7 \in \frac{1}{448} {\cal Q}_{16}$\\
& $x_8\in\{\pm \frac{1}{8}j \}$& $x_8\in\frac{1}{128}{\cal Q}_4$& $ x_8 \in \frac{1}{896} {\cal Q}_{16}$\\
& $x_9\in\{\pm \frac{1}{16}\}$ & $x_9\in\frac{1}{256}{\cal Q}_4$& $ x_9 \in \frac{1}{1792} {\cal Q}_{16}$\\
& $x_{10}\in\{\pm \frac{1}{16}j\}$& $x_{10}\in\frac{1}{512}{\cal Q}_4$ & $ x_{10} \in \frac{1}{3584} {\cal Q}_{16}$\\
& $x_{11}\in\{\pm \frac{1}{32} \}$& $x_{11}\in\frac{1}{1024}{\cal Q}_4$ & $ x_{11} \in \frac{1}{7168} {\cal Q}_{16}$ \\
& $x_{12}\in\{\pm \frac{1}{32}j \}$& $x_{12}\in\frac{1}{2048}{\cal Q}_4$ & $ x_{12} \in \frac{1}{14336}{\cal Q}_{16}$\\
& $x_{13}\in\{\pm \frac{1}{64}\}$& $x_{13}\in\frac{1}{4096}{\cal Q}_4$& $ x_{13} \in \frac{1}{28672} {\cal Q}_{16}$\\
& $x_{14}\in\{\pm \frac{1}{64}j\}$& $x_{14}\in\frac{1}{8192}{\cal Q}_4$ & $ x_{14} \in \frac{1}{57344} {\cal Q}_{16}$\\
& $x_{15}\in\{\pm\frac{1}{128}\}$& $x_{15}\in\frac{1}{16384}{\cal Q}_4$ & $ x_{15} \in \frac{1}{114688} {\cal Q}_{16}$\\
& $x_{16}\in\{\pm\frac{1}{128}j \}$& $x_{16}\in\frac{1}{32768}{\cal Q}_4$& $ x_{16} \in \frac{1}{229376} {\cal Q}_{16}$\\
\hline
\end{tabular}
\caption{Full diversity achieving constellation sets for $n_t=3,4,8,16$
with 1 bit/symbol, 2 bits/symbol and 3 bits/symbol obtained by computer 
search.}
\end{center}
\end{table}

\vspace{-4mm}
\section{Results and Discussions}
\label{simu}
\vspace{-4mm}
We evaluated the codeword error rate (CER) performance of the proposed
precoding scheme in asymmetric MIMO systems with $n_r<n_t$ through 
simulations. In particular, we illustrate the CER performance of the
proposed scheme for $3\times 1$, $3\times 2$, $4 \times 1$ 
and $8\times 1$ full-rate MIMO as a function of the average 
received SNR per receive antenna. ML decoding is used. 

\begin{figure}
\centering
\epsfxsize=16.25cm
\epsfysize=9.5cm
\epsfbox{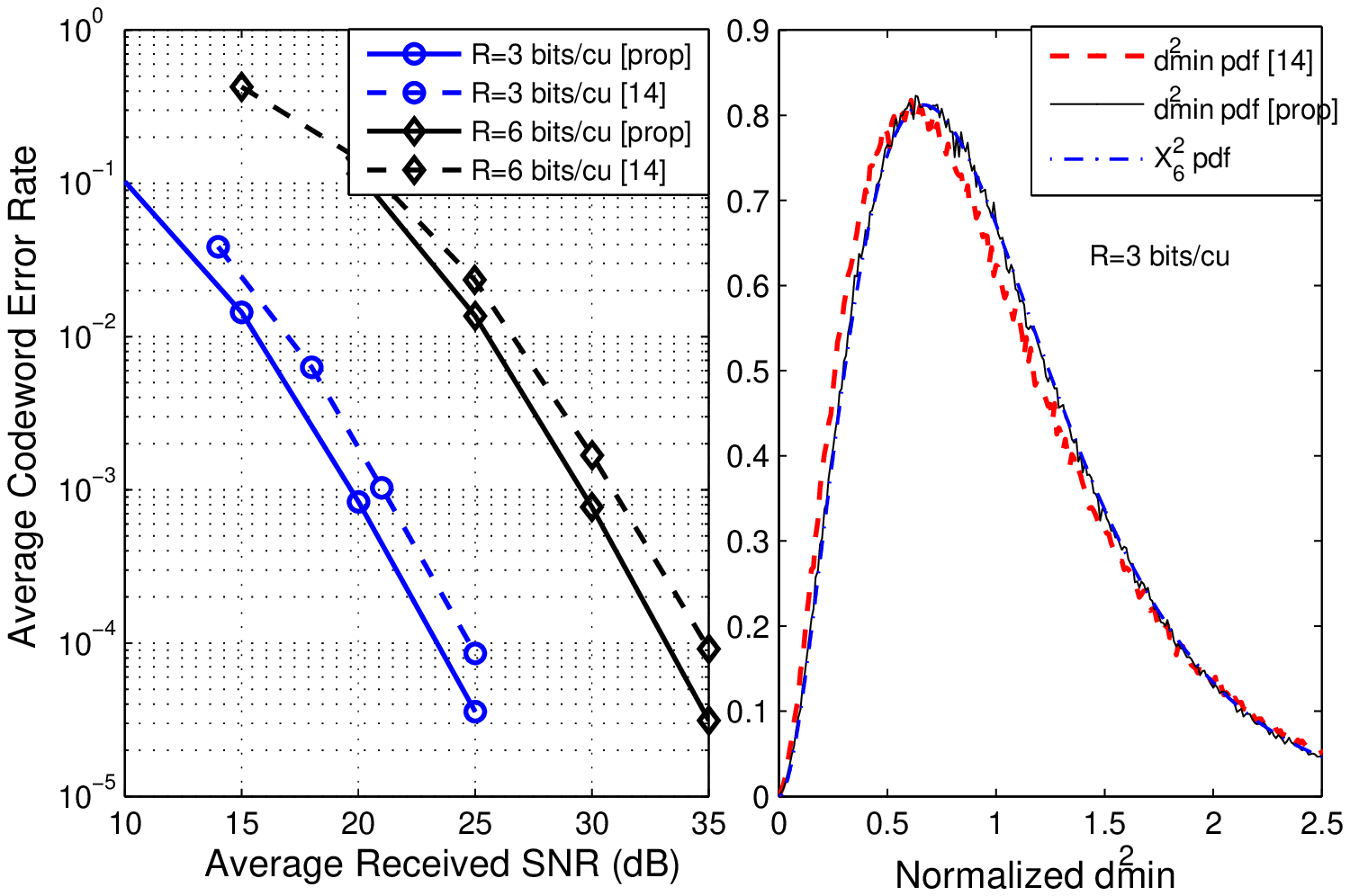}
\caption{Comparison of $a)$ CER performance and $b)$ pdf of the
normalized $d_{min}^2$ of the proposed precoder with that in \cite{BARIK}
for $3\times 1$ MIMO system.}
\label{fig_2x1}
\end{figure}

\begin{figure}
\centering
\epsfxsize=16.00cm
\epsfysize=9.5cm
\epsfbox{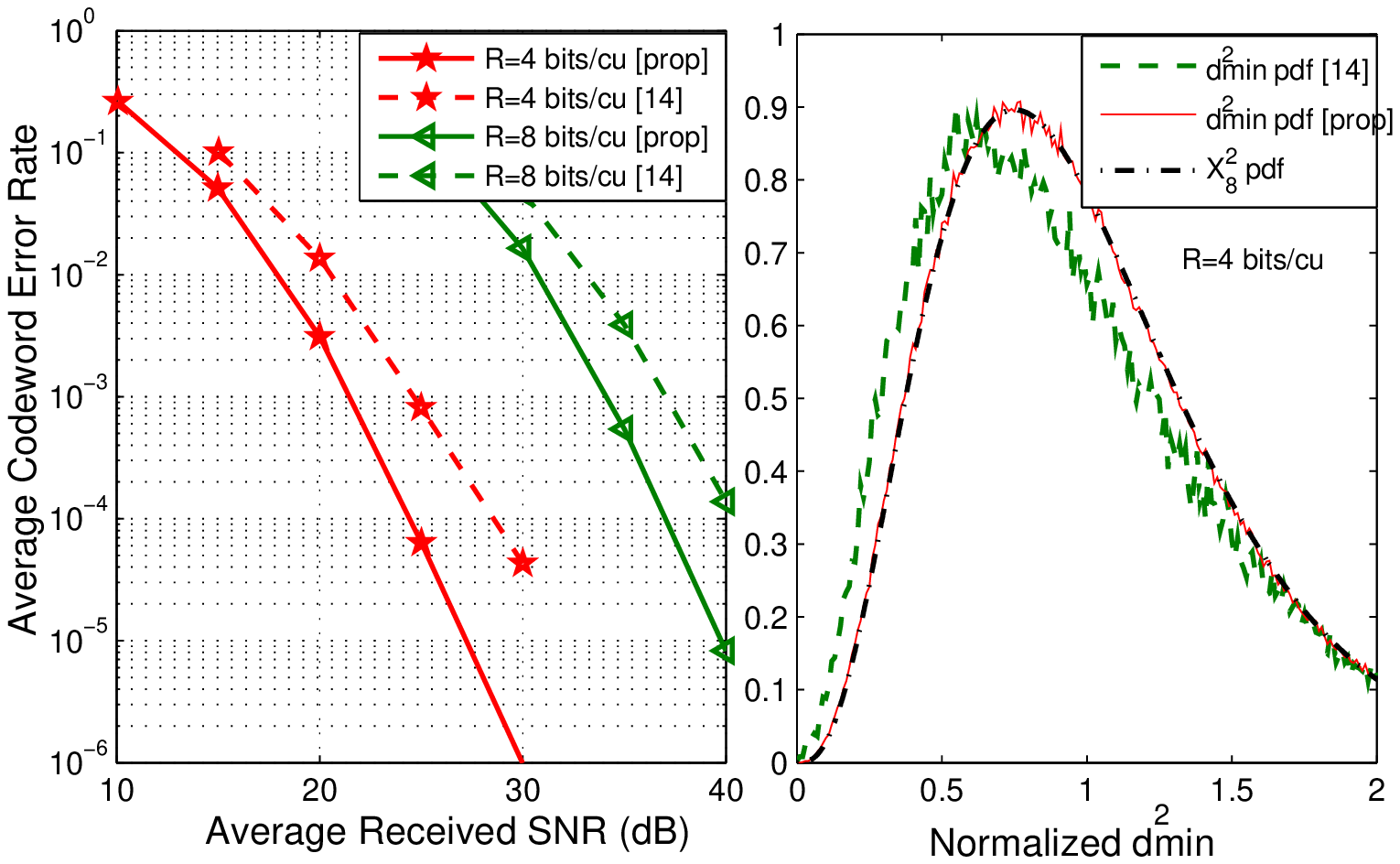}
\caption{Comparison of $a)$ CER performance and $b)$ pdf of the
normalized $d_{min}^2$ of the proposed precoder with that in \cite{BARIK}
for $4\times 1$ MIMO system. }
\label{fig_3x1}
\end{figure}

\begin{figure}
\centering
\epsfxsize=16.25cm
\epsfysize=9.5cm
\epsfbox{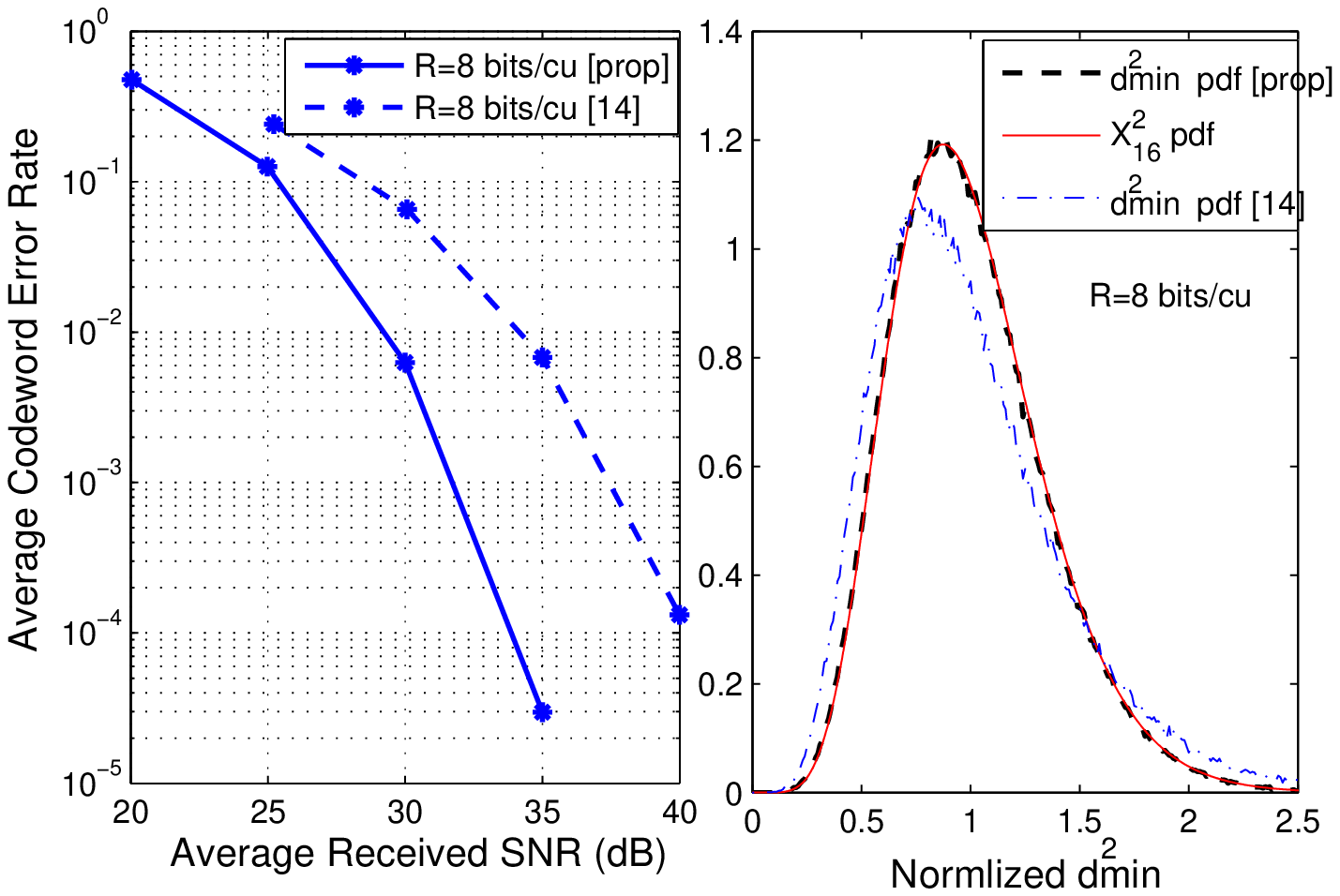}
\caption{Comparison of $a)$ CER performance and $b)$ pdf of the
normalized $d_{min}^2$ of the proposed precoder with that in \cite{BARIK}
for $8\times 1$ MIMO system.}
\label{fig_4x1}
\end{figure}

Plots of two performance measures, namely, $i)$ CER and $ii$) pdf of the 
normalized $d_{min}^2$, are shown in two separate sub-figures in each 
of Figs. \ref{fig_2x1}, \ref{fig_3x1}, \ref{fig_4x1} for $3\times 1$, 
$4\times 1$ and $8\times 1$ MIMO, respectively. For comparison purposes, 
we have plotted the CER performance and $d_{min}^2$ pdf of the full-rate 
precoding scheme in \cite{BARIK}. We show the comparison only with 
\cite{BARIK} because both the proposed scheme and the scheme in 
\cite{BARIK} achieve the full rate of $n_t$ symbols per channel use 
in $n_r<n_t$ asymmetric MIMO. Whereas, the other precoding schemes in 
Table-1 lose rate and do not achieve full rate of $n_t$ symbols per 
channel use in $n_r<n_t$ asymmetric MIMO.

In Fig. \ref{fig_2x1}, CER plots for the rates of 3 and 6 bits/channel
use are shown. For 3 bits/cu, the constellation points used in the
simulation for the proposed scheme are as given in the entries of
$n_t=3$ and 1 bit/symbol modulation in Table-2. Likewise, for 6 bits/cu,
the constellation points used are from the $n_t=3$ and 2 bits/symbol
modulation in Table-2. Appropriate constellations from Table-2 are used
for the plots of different rates shown in Figs. \ref{fig_3x1} and
\ref{fig_4x1}. Since constellation optimization is not done in the 
precoding scheme in \cite{BARIK}, we have used regular modulation 
alphabets (e.g., BPSK, 4-QAM) and matched the bits/cu in both schemes 
for fair comparison. For example, 6 bits/cu plot in Fig. \ref{fig_2x1} 
for the scheme in \cite{BARIK} is simulated using 4-QAM.

\begin{figure}
\centering
\epsfxsize=16.25cm
\epsfysize=9.5cm
\epsfbox{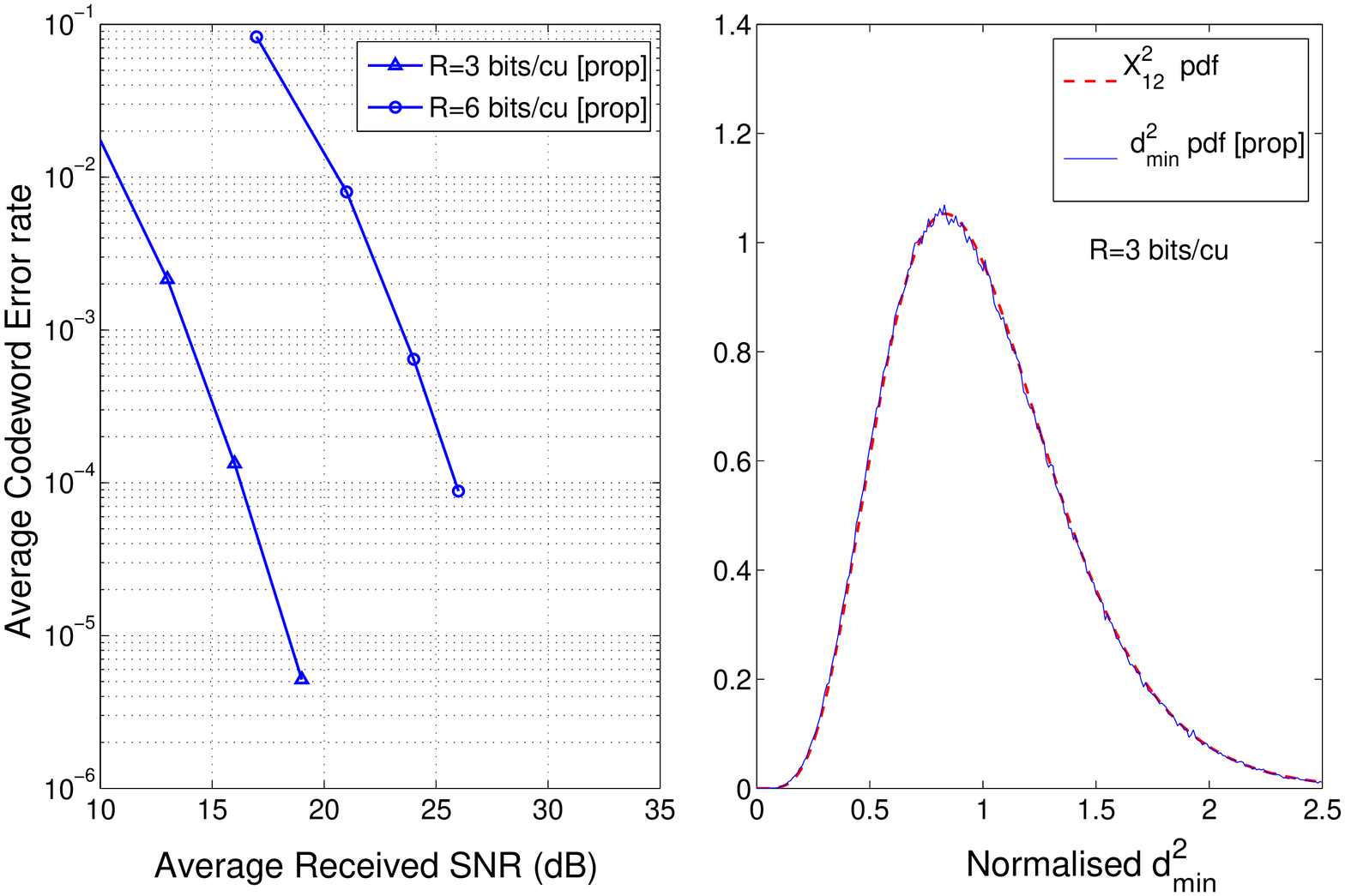}
\caption{CER performance and pdf of the normalized $d_{min}^2$ of the
proposed precoder in a $3\times 2$ MIMO system.}
\label{fig_3x2}
\end{figure}

From the CER plots of in Figs. \ref{fig_2x1} to \ref{fig_4x1}, we can
see that the proposed precoding achieves better diversity orders compared
to the precoding scheme in \cite{BARIK}. Indeed, as predicted by the
analysis, the proposed scheme
achieves the full diversity of $n_tn_r= 3,4,8$ in Figs. \ref{fig_2x1},
\ref{fig_3x1} and \ref{fig_4x1}, respectively. This can be verified by
observing that pdfs of the $d^2_{min}$ in the proposed scheme match
with the theoretical $\chi_{2n_tn_r}^2$ pdf (Chi square distribution with
$2n_tn_r$ degrees of freedom). For example, in Fig. \ref{fig_3x1}, the 
proposed scheme's pdf matches with that of the $\chi_{8}^2$ pdf, verifying 
the achievability of $n_tn_r=4$th order diversity. Similar matches with 
Chi square distribution are observed in other figures as well. Thus,
the analytical claim of full diversity of $n_tn_r$ in the proposed
precoding scheme is validated through simulations as well. It is further
noted that the pdfs in the scheme in \cite{BARIK} do not match with
those of the theoretical $\chi_{2n_tn_r}^2$ pdfs, indicating that the
scheme in \cite{BARIK} does not achieve full diversity. Finally, Fig.
\ref{fig_3x2} shows the performance of the proposed scheme in a
$3\times 2$ MIMO system with 3 and 6 bits/cu, where the full $n_tn_r=6$th
order diversity is found to be achieved.

\vspace{-4mm}
\section{Conclusion}
\label{sec6}
\vspace{-4mm}
We presented a {\em partial CSIT} based precoding scheme, which achieved
both {\em full diversity} ($n_tn_r$th diversity) as well as {\em full rate}
($n_t$ symbols per channel use) with a feedback requirement of only
$n_t-1 $ real angular parameters, applicable in MIMO systems including 
asymmetric MIMO with $n_r<n_t$. The full diversity was achieved by
choosing optimized constellation sets. Through analysis and simulations
we established the full diversity achievability in the proposed scheme.

\newpage

\end{document}